\documentclass[preprint,prd,aps,showpacs,showkeys,nofootinbib]{revtex4}
\usepackage{graphicx}
\usepackage{color}
\begin{document}
\title{Study of dark matter in the extended BLMSSM}
\author{Jing-Jing Feng$^{1}$\footnote{fengjj$\_$0726@163.com},
Shu-Min Zhao$^{1}$\footnote{zhaosm@hbu.edu.cn},
Xing-Xing Dong$^{1}$,
Zhong-Jun Yang$^{1}$,
Hai-Bin Zhang$^{1}$,
Fang Wang$^{2}$,
Tai-Fu Feng$^{1}$\footnote{fengtf@hbu.edu.cn}}

\affiliation{&&$^1$ Department of Physics and Technology, Hebei University, Baoding 071002, China
\nonumber\\$^{2}$ College of Electronic Information Engineering, Hebei University, Baoding 071002, China}
\begin{abstract}
There are strong evidences for existence of dark matter in some experiments at present. However, the question is that we do not have a reasonable explanation for dark matter in the framework of the Standard Model(SM) of particle physics. It is necessary to extend the SM in order to explain the dark matter. According to the current possible existence conditions of dark matter, we choose $\chi^0_L$ and $\tilde{Y}$ as candidates for dark matter in the EBLMSSM. We study the dominant annihilation processes in detail, including $\bar{\chi}^0_L\chi^0_L(\bar{\tilde{Y}}\tilde{Y})\rightarrow \bar{l}^Il^I$ and
$\bar{\chi}^0_L\chi^0_L(\bar{\tilde{Y}}\tilde{Y})\rightarrow \bar{\nu}^I\nu^I$. And we calculate their annihilation cross section $\sigma$ and relic density $\Omega_D h^2$. Then we analyze the limitations of dark matter relic density on the parameters of  the EBLMSSM.
\end{abstract}

\maketitle
\section{Introduction}
Astronomers are convinced that there are an amazing amount of dark matter in the universe by some astronomical observations and theoretical derivations.
The earliest and perhaps the most convincing evidence of the existence of dark matter until today comes from observations of the rotation curves of
galaxies\cite{rotation curve of galaxy1}, namely the graph of circular velocities of stars and gas as a function of their distance from the galactic center.
This observation is inconsistent with our calculations using Newtonian dynamics. Therefore, it can be inferred that there are a lot of invisible matter in
the universe, which we call dark matter(DM)\cite{rotation curve of galaxy2, name, rotation curve of galaxy3}. If you want to know more evidences of the existence
of dark matter, you can find them in Ref\cite{other exist1, other exist2, other exist3, other exist4, other exist5}. Dark matter is widespread and abundant,
and it accounts for about 23\% of the universe, while the common baryon matter only accounts for about 4\%\cite{account1, account2}. However, the standard model(SM) of the particle physics can not provide a convincing explanation. So studying dark matter is a meaningful and interesting work to explore new physics beyond SM.

Although we don't know what dark matter is and how it exists, we can deduce some of its properties by analyzing and calculating data. First, dark matter particles have to be both electrically and color neutral. Second, it only participates in weak interactions. Finally it must remain stable or have a long life, otherwise it will decay into other particles\cite{characteristic1, characteristic2}. Only neutrinos in SM can have the correct interaction properties, but it is now known that the masses of neutrinos are too small to constitute the main component of dark matter. Based on the above conditions, there are no suitable dark matter candidates in SM. What is certain is that the existence of suitable dark matter particles is beyond SM. In recent years, weakly-interacting massive particles(WIMPS) have become the most popular candidates for cold dark matter. Many experiments are detecting it, so experimental limitations have been strengthening.

In fact, not only the problem of dark matter but also many other phenomena are not explained well by SM, and these problems and phenomena may become strong evidences for the exploration of new physics. The minimal supersymmetric extension of the standard model(MSSM) is a popular theory to explain these anomalies. And physicists in various countries have been studying it for many years. BLMSSM where the baryon and lepton numbers are local gauge symmetries spontaneously broken at the TeV scale is the simple extension of MSSM. Since the matter-antimatter asymmetry in the universe naturally, the baryon number(B) is broken. On the other hand, considering the neutrino oscillation experiment, the heavy majorana neutrinos contained in the seesaw mechanism can induce the tiny neutrino masses, therefore, the lepton number(L) is also expected to be broken. The proton remains stable and R-parity is not conserved. Consequently the predictions and bounds for the collider experiments should be changed in the BLMSSM.

Although BLMSSM can explain many anomalies, we find that the exotic leptons in BLMSSM is not heavy enough. The diagonal elements of their mass matrix is zero. Therefore, the masses of the exotic leptons are only related to the four parameters $Y_{e4}$, $Y_{e5}$, $\upsilon_d$, and $\upsilon_u$. Here $\upsilon_d$ and $\upsilon_u$ are the vacuum expectation values(VEVs) of two Higgs doublets $H_d$ and $H_u$. They need to satisfy the equation $\sqrt{\upsilon^2_d+\upsilon^2_u}=\upsilon \sim 250$GeV. At the same time, the Yukawa couplings $Y_{e4}$ and $Y_{e5}$ are not large parameters. It can be calculated that the masses of the exotic leptons are approximately 100GeV. With the advancement of high energy physics experiments, this boundary will soon be ruled out in the future. It is related to whether the BLMSSM can continue to exist. Therefore, in order to get heavy exotic leptons, we add two exotic Higgs superfields which are $SU(2)$ singlets $\Phi_{NL}$ and $\varphi_{NL}$ with VEVs of $\upsilon_{NL}$ and $\bar{\upsilon}_{NL}$ to the BLMSSM. In this way, the mass matrix of the exotic leptons become Eq.(\ref{exotic lepton mass matrix}). These diagonal elements can be large and contribute to mass, so the masses of exotic leptons become heavy enough. Because the heavy exotic leptons should be unstable, the two superfields $Y$ and $Y^{'}$ are introduced. On the other hand, the fourth and fifth generation leptons are mixed, which is different from the BLMSSM. It is obvious that the first four terms of $\mathcal{W}_B$ and $\mathcal{W}_L$ in Eq.(1) are exactly corresponding. We call this new model EBLMSSM which is an extending of BLMSSM.
Fortunately, we also find several new candidates for cold dark matter in the EBLMSSM. In our previous work, we have studied $Y$ as dark matter candidate in the EBLMSSM \cite{Y of zhao}and we study $\chi^0_L$ and $\tilde{Y}$ in this paper. Here, we study dark matter annihilating into leptons and light neutrinos in the EBLMSSM.

The biggest clue is the observation of the relic density for dark matter in the universe, which is a strong constraint on the model to explain dark matter until now. The latest experimental observations show that the dark matter relic density is 0.1186$\pm$0.0020\cite{numerical}. Furthermore, there are constraints on the mass of Higgs. In general, the EBLMSSM meet the above constraints. In our work, the DM relic density should satisfy 0.1186$\pm$0.0020 within 3$\sigma$ range. It is strictly limited, which results in the parameters in the EBLMSSM to vary only in a narrow range. Our work is only to provide a possibility to explain dark matter, and also to provide a direction for indirect detection experiments.

The remaining of the paper is organized as follows: In section II we introduce the EBLMSSM model in detail. In section III we discuss the relic density in the universe at present and we calculate annihilation section of dark matter candidates. Section IV is focused on the numerical analysis. In section V, we give our conclusions.

\section{introduction of the Model}
In this section, we briefly introduce the basic characteristics of EBLMSSM. It is the extension of BLMSSM. The local gauge group is
$SU(3)_{C}\otimes SU(2)_{L}\otimes U(1)_{Y}\otimes U(1)_{B}\otimes U(1)_{L}$\cite{group1, group2, group3, group4}. Compared with the BLMSSM, the EBLMSSM includes
 four new superfields and some new particles. In order to generate large mass for the exotic leptons, we need to introduce the two new superfields
 ($\Phi_{NL}$ and $\varphi_{NL}$) with nonzero vacuum expectation values ( $\upsilon_{NL}$ and $\bar{\upsilon}_{NL}$). At the same time, the other two new superfields ($Y$ and $Y'$) also are added to keep the heavy exotic leptons unstable\cite{Y of zhao}.
The superpotential of EBLMSSM is given by
\begin{eqnarray}
&&{\cal W}_{{EBLMSSM}}={\cal W}_{{MSSM}}+{\cal W}_{B}+{\cal W}_{L}+{\cal W}_{X}+{\cal W}_{Y}\;,
\nonumber\\
&&{\cal W}_{B}=\lambda_{Q}\hat{Q}_{4}\hat{Q}_{5}^c\hat{\varphi}_{B}+\lambda_{U}\hat{U}_{4}^c\hat{U}_{5}
\hat{\Phi}_{B}+\lambda_{D}\hat{D}_{4}^c\hat{D}_{5}\hat{\Phi}_{B}
+\mu_{B}\hat{\Phi}_{B}\hat{\varphi}_{B}\nonumber\\&&\hspace{1.2cm}+Y_{{u_4}}\hat{Q}_{4}\hat{H}_{u}\hat{U}_{4}^c+Y_{{d_4}}\hat{Q}_{4}\hat{H}_{d}\hat{D}_{4}^c
+Y_{{u_5}}\hat{Q}_{5}^c\hat{H}_{d}\hat{U}_{5}+Y_{{d_5}}\hat{Q}_{5}^c\hat{H}_{u}\hat{D}_{5}\;,
\nonumber\\
&&{\cal W}_{L}=\lambda_{L}\hat{L}_{4}\hat{L}_{5}^c\hat{\varphi}_{NL}+\lambda_{E}\hat{E}_{4}^c\hat{E}_{5}
\hat{\Phi}_{NL}+\lambda_{NL}\hat{N}_{4}^c\hat{N}_{5}\hat{\Phi}_{NL}
+\mu_{NL}\hat{\Phi}_{NL}\hat{\varphi}_{NL}\nonumber\\&&\hspace{1.2cm}+Y_{{e_4}}\hat{L}_{4}\hat{H}_{d}\hat{E}_{4}^c+Y_{{\nu_4}}\hat{L}_{4}\hat{H}_{u}\hat{N}_{4}^c
+Y_{{e_5}}\hat{L}_{5}^c\hat{H}_{u}\hat{E}_{5}+Y_{{\nu_5}}\hat{L}_{5}^c\hat{H}_{d}\hat{N}_{5}
\nonumber\\
&&\hspace{1.2cm}
+Y_{\nu}\hat{L}\hat{H}_{u}\hat{N}^c+\lambda_{{N^c}}\hat{N}^c\hat{N}^c\hat{\varphi}_{L}
+\mu_{L}\hat{\Phi}_{L}\hat{\varphi}_{L}\;,
\nonumber\\&&
{\cal W}_{Y}=\lambda_4\hat{L}\hat{L}_{5}^c\hat{Y}+\lambda_5\hat{N}^c\hat{N}_{5}\hat{Y}^\prime
+\lambda_6\hat{E}^c\hat{E}_{5}\hat{Y}^\prime+\mu_{Y}\hat{Y}\hat{Y}^\prime\;.\label{super}
\end{eqnarray}
where ${\cal W}_{{MSSM}}$ represents the superpotential of the MSSM. ${\cal W}_{B}$ and ${\cal W}_{X}$ are same as the terms in the BLMSSM. ${\cal W}_{L}$ is different from BLMSSM for adding the first four items in Eq.(\ref{super})\cite{model1, model2}. ${\cal W}_{Y}$ has some new couplings including the lepton-exotic lepton-$Y$ coupling and lepton-exotic slepton-$\tilde{Y}$ coupling. Furthermore, we can also acquire lepton-slepton-lepton neutralino coupling. Here $\tilde{Y}$ is the superpartners of $Y$ and $Y^{'}$ and it's four component spinor. In fact, The new couplings of $l^I-L^{'}-Y$ and $l^I-\tilde{L}^{'}-\tilde{Y}$ have a great influence on lepton anormal magnetic dipole moment(MDM) in one loop order. They are able to correct the muon MDM and match the experimental values very well. New parameter $\mu_{Y}$ provides a new source for CP-violating. If $\lambda_4(\lambda_6)$ has non-zero elements about lepton flavor, $\mathcal{W}_Y$ can enhance the impact of lepton flavor violating\cite{worth}. In short, these new couplings and new parameters enrich the lepton physics to a certain degree. In addition, study of dark matter has been promoted. It provides a new possibility for explaining dark matter. Besides the above mentioned problems, we can also study many other new physics problems in the EBLMSSM. Of course, these are our future work. There are one Majorana fermion($\chi^0_L$), two Dirac fermions ($\tilde{Y}$ and $\tilde{X}$), and two scalar particles ($Y$ and $X$) as good dark matter candidates in EBLMSSM. Among them, three particles($\tilde{X}$, $X$ and $Y$) has been discussed in previous work\cite{group2, Y of zhao}. The other particles($X^0_L$ and $\tilde{Y}$) will be discussed in this paper.

Based on the new introduced superfields $\Phi_{NL},\varphi_{NL}, Y$ and $Y'$ in the EBLMSSM, the soft breaking terms are as follows
\begin{eqnarray}
&&{\cal L}_{{soft}}^{EBLMSSM}={\cal L}_{{soft}}^{BLMSSM}
-m_{{\Phi_{NL}}}^2\Phi_{NL}^*\Phi_{NL}
-m_{{\varphi_{NL}}}^2\varphi_{NL}^*\varphi_{NL}
+(A_{{LL}}\lambda_{L}\tilde{L}_{4}\tilde{L}_{5}^c\varphi_{NL}
\nonumber\\&&\hspace{2.5cm}
+A_{{LE}}\lambda_{E}\tilde{e}_{4}^c\tilde{e}_{5}\Phi_{NL}
+A_{{LN}}\lambda_{NL}\tilde{\nu}_{4}^c\tilde{\nu}_{5}\Phi_{NL}
+B_{NL}\mu_{NL}\Phi_{NL}\varphi_{NL}+h.c.)
\nonumber\\&&\hspace{2.5cm}+(
A_4\lambda_4\tilde{L}\tilde{L}_{5}^cY+A_5\lambda_5\tilde{N}^c\tilde{\nu}_{5}Y^\prime
+A_6\lambda_6\tilde{e}^c\tilde{e}_{5}Y^\prime+B_{Y}\mu_{Y}YY^\prime+h.c.).
\label{soft-breaking}
\end{eqnarray}
${\cal L}_{{soft}}^{BLMSSM}$ is the soft breaking terms of the BLMSSM discussed in our previous work. $SU(2)_L$ singlets $\Phi_{L}$ , $\varphi_{L}$ , $\Phi_{NL}$ ,$\varphi_{NL}$ acquire the nonzero VEVs $\upsilon_{L} $ , $\bar{\upsilon}_{L}$ , $\upsilon_{NL} $ , $\bar{\upsilon}_{NL}$ respectively.
The local gauge symmetry $SU(2)_{L}\otimes U(1)_{Y}\otimes U(1)_{B}\otimes U(1)_{L}$ breaks down to electromagnetic symmetry $U(1)_{e}$,
\begin{eqnarray}
&&\Phi_{L}={1\over\sqrt{2}}\Big(\upsilon_{L}+\Phi_{L}^0+iP_{L}^0\Big),~~~~
\varphi_{L}={1\over\sqrt{2}}\Big(\bar{\upsilon}_{L}+\varphi_{L}^0+i\bar{P}_{L}^0\Big),
\nonumber\\&&
\Phi_{NL}={1\over\sqrt{2}}\Big(\upsilon_{NL}+\Phi_{NL}^0+iP_{NL}^0\Big),~~~~
\varphi_{NL}={1\over\sqrt{2}}\Big(\bar{\upsilon}_{NL}+\varphi_{NL}^0+i\bar{P}_{NL}^0\Big).
\label{singlets}
\end{eqnarray}

In the EBLMSSM, some mass matrices are different from BLMSSM because of the introduced superfields $\Phi_{NL}$ and $\varphi_{NL}$. We list some mass matrices and new couplings as following. If you want to know more, whether it is the mass matrix or the coupling, you can find it in our previous work\cite{matrix and coupling1, matrix and coupling2}.
\subsection{the mass matrices}
\subsubsection{The lepton neutralino mass matrix in the EBLMSSM}
In the EBLMSSM, $\lambda_L$, the superpartner
of the new lepton type gauge boson $Z^\mu_L$, mixes with the SUSY
superpartners $(\psi_{\Phi_L},\psi_{\varphi_L},\psi_{\Phi_{NL}},\psi_{\varphi_{NL}})$ of the superfields ($\Phi_{L},\varphi_{L},\Phi_{NL},\varphi_{NL}$). So the lepton neutralino mass matrix is obtained in the base $(i\lambda_L,\psi_{\Phi_L},\psi_{\varphi_L},\psi_{\Phi_{NL}},\psi_{\varphi_{NL}})$,
\begin{equation}
\mathcal{M}_L=\left(     \begin{array}{ccccc}
  2M_L &2\upsilon_Lg_L &-2\bar{\upsilon}_Lg_L&3\upsilon_{NL}g_L &-3\bar{\upsilon}_{NL}g_L\\
   2\upsilon_Lg_L & 0 &-\mu_L& 0 & 0\\
   -2\bar{\upsilon}_Lg_L&-\mu_L &0& 0 & 0\\
   3\upsilon_{NL}g_L & 0 & 0 & 0 & -\mu_{NL}\\
   -3\bar{\upsilon}_{NL}g_L& 0&0&-\mu_{NL}&0
    \end{array}\right).
\label{X mass matrix1}
\end{equation}
The mass matrix $\mathcal{M}_L$ can be diagonalized by the rotation matrix $Z_{NL}$. Then, we can have
\begin{eqnarray}
&&i\lambda_L=Z_{NL}^{1i}K_{L_i}^0
,~~~\psi_{\Phi_L}=Z_{NL}^{2i}K_{L_i}^0
,~~~\psi_{\varphi_L}=Z_{NL}^{3i}K_{L_i}^0,
\nonumber\\&&\psi_{\Phi_{NL}}=Z_{NL}^{4i}K_{L_i}^0
,~~~~~\psi_{\varphi_{NL}}=Z_{NL}^{5i}K_{L_i}^0.
\label{X mass matrix2}
\end{eqnarray}
Here, $X^0_{L_i}=(K_{L_i}^0,\bar{K}_{L_i}^0)^T$ represent the mass egeinstates of the lepton neutralino.
\subsubsection{The exotic lepton mass matrix in the EBLMSSM}
The mass matrix for the exotic leptons reads as
\begin{eqnarray}
&&-\mathcal{L}^{mass}_{e^{'}}=(\bar{e}^{'}_{4R},\bar{e}^{'}_{5R}) \left(  \begin{array}{ccccc}
  -\frac{1}{\sqrt{2}}\lambda_L \bar{\upsilon}_L &-\frac{1}{\sqrt{2}}Y_{e5}\upsilon_u\\
  -\frac{1}{\sqrt{2}}Y_{e4}\upsilon_d &\frac{1}{\sqrt{2}}\lambda_{E}\upsilon_L
\end{array}\right)\left( \begin{array}{c}
  e^{'}_{4L} \\   e^{'}_{5L}\\
    \end{array}\right)+h.c.
\label{exotic lepton mass matrix}
\end{eqnarray}
\subsubsection{The $\tilde{Y}$ mass matrix in the EBLMSSM}
The mass term for superfield $\tilde{Y}$ in the Lagrangian is given out
\begin{eqnarray}
  &&-\mathcal{L}^{mass}_{\tilde{Y}}=\mu_Y\bar{\tilde{Y}}\tilde{Y}
  ,~~~~~~~~~~~~~~~~\tilde{Y} =\left( \begin{array}{c}
  \psi_{Y'} \\  \bar{\psi}_{Y}\\
    \end{array}\right).
\label{Y mass matrix}
\end{eqnarray}

Here $m_{\tilde{Y}}$ (the mass of $\tilde{Y}$)=$\mu_Y$.
\subsection{some couplings}
\subsubsection{The couplings with  $\chi^0_L$}
As dark matter candidate, lepton neutralinos($\chi^0_L$) not only have relations with leptons($l$) and sleptons($\tilde{L}$), but also act with neutrinos($\nu$) and sneutrinos($\tilde{\nu}$).
\begin{eqnarray}
&&\mathcal{L}(\chi^0_L\tilde{L}l)=\sum^3_{I=1}\sum^6_{i=1}\sqrt{2}g_L\bar{\chi}^0_{L_j}(Z^{1j}_{NL}Z^{Ii}_{\tilde{L}}P_L-Z^{1j*}_{NL}Z^{(I+3)i}_{\tilde{L}}P_R)l^I\tilde{L}^+_i+h.c.
\label{coupling1}
\end{eqnarray}
\begin{eqnarray}
&&\mathcal{L}(\chi^0_L\tilde{\nu}\nu)=\sum^3_{I,J=1}\sum^6_{\alpha,j=1}\bar{X}_{N_\alpha}([-(\lambda^{IJ}_{N_c}+\lambda^{JI}_{N_c})Z^{(I+3)\alpha}_{N_\nu}Z^{3i}_{N_L}Z^{(J+3)j*}_{\tilde{\nu}}+\sqrt{2}g_LZ^{I\alpha}_{N_\nu}Z^{1i}_{N_L}Z^{Jj*}_{\tilde{\nu}}\delta_{IJ}]P_L
\nonumber\\&&\hspace{2.2cm}-\sqrt{2}g_LZ^{(I+3)\alpha*}_{N_\nu}Z^{1i*}_{N_L}Z^{(J+3)j*}_{\tilde{\nu}}\delta_{IJ}]P_R)X^0_{L_i}\tilde{\nu}^{j*}+h.c.
\label{coupling2}
\end{eqnarray}
 The couplings for lepton neutralino-new gauge boson $Z^{\mu}_L$-lepton neutralino read as
\begin{eqnarray}
&&\mathcal{L}(X^0_LX^0_LZ^\mu_L)=-g_LZ^\mu_L\bar{X}^0_{L_i}\gamma_\mu(3Z^{5i*}_{N_L}Z^{5j}_{N_L}-3Z^{4i*}_{N_L}Z^{4j}_{N_L})P_LX^0_{L_j}+h.c.
\label{coupling3}
\end{eqnarray}

\subsubsection{The couplings with  $\tilde{Y}$}
For  another dark matter candidate $\tilde{Y}$, in addition to the interactions with exotic sleptons($\tilde{L'}$) and leptons($l$), there are also interactions with exotic neutrinos($\tilde{N'}$) and neutrinos($\nu$), whose couplings are in the following form
\begin{eqnarray}
&&\mathcal{L}(\tilde{Y}l\tilde{L'})=\sum^3_{I=1}\sum^4_{i=1}\bar{\tilde{Y}}(-\lambda_4Z^{4i*}_{\tilde{L'}}P_L-\lambda_6Z^{3i*}_{\tilde{L'}}P_R)e^I\tilde{L'}^*_i+h.c.
\label{coupling4}
\end{eqnarray}
\begin{eqnarray}
&&\mathcal{L}(\tilde{Y}\nu\tilde{N'})=\sum^3_{I=1}\sum^4_{i=1}\sum^6_{\alpha=1}\bar{\tilde{Y}}(-\lambda_4Z^{I\alpha}_{N_\nu}Z^{4i*}_{\tilde{N'}}P_L-\lambda_5Z^{(I+3)\alpha*}_{N_\nu}Z^{3i*}_{\tilde{N'}}P_R)X^0_{N_\alpha}\tilde{N}^{'*}_i+h.c.
\label{coupling5}
\end{eqnarray}

$\tilde{Y}$ interacts with $Z^{\mu}_L$ and $\tilde{Y}$, whose coupling is in the following form
\begin{eqnarray}
&&\mathcal{L}(\tilde{Y}\tilde{Y}Z^{\mu}_L)=\bar{\tilde{Y}}[g_L(2+L_4)\gamma_\mu]Z^\mu_L\tilde{Y}+h.c.
\label{coupling6}
\end{eqnarray}

The new gauge boson $Z^{\mu}_L$ couples with leptons and neutrinos, whose couplings can be find in our previous work\cite{Y of zhao}.

\section{Dark matter candidates: $X^0_L$ and $\tilde{Y}$}
 In this section, we suppose the lightest mass eigenstate of $X^0_L$ and $\tilde{Y}$ as dark matter candidates. And they belong to the scope of weakly-interacting massive particles(WIMPS) that are the most studied dark matter candidates. WIMPS have masses in the range 10GeV-TeV and tree level interactions with the W and Z gauge bosons, but not with photons\cite{rotation curve of galaxy1, WIMPS1, WIMPS2, WIMPS3}. So we summarize the relic density contraint that any WIMP candidates are satified. First, we need to introduce the freeze-out temperature $T_F$. And $T_F$ is usually expressed as a dimensionless quantity $x=\frac{m_D}{T}$ when $T_F=T$, where $m_D$ is dark matter mass. Then, we get the concrete form of $x_F$ by solving the Boltzmann equation of the dark matter number density n\cite{boltzmann1, boltzmann2, boltzmann3, boltzmann4}.
 \begin{eqnarray}
 &&\dot{n}=-3Hn-\langle \sigma v \rangle(n^2-n^2_0).\label{jiemian1}
\end{eqnarray}

Here $n$ is the number density of the dark matter, $\sigma$ is the annihilation cross section of the particle, $v$ is the relative velocity of the annihilating particles, $H$ is the expansion rate of the Universe and $n_0$ is the dark matter number density in thermally equilibrium. Finally, we can obtain an iterative equation about $x_F$ for solving Eq.(\ref{jiemian1}).
 \begin{eqnarray}
 &&x_F=\ln\frac{0.076 M_{pl} m_D \langle\sigma v\rangle}{\sqrt{g_{\ast} x_F}}.
\label{jiemian2}
\end{eqnarray}
Here, $M_{pl}=1.22\times 10^{19}$GeV is the Planck mass and $g_*$ is the number of the relativistic degrees of freedom with mass less
than $T_F$. We can calculate cross section and the term $\langle \sigma v \rangle$ in the Eqs.(\ref{jiemian1}-\ref{jiemian2}) can be written as
\begin{eqnarray}
 &&\langle \sigma v \rangle=a+b v^2+\mathcal{O}(v^4).
\label{jiemian3}
\end{eqnarray}
Notice $a$ and $b$ are the first two coefficients in the Taylor expansion of the annihilation cross section\cite{Y of zhao, boltzmann2,ab}.
We begin with a brief calculation formula of the present relic density($\Omega_D h^2$) of DM candidates, assuming that the mass $m_D$ as well as the annihilation cross section $\sigma$ are known. Furthermore, neglecting terms which are $\mathcal{O}(v^4)$, we give the expression of $x_F$ and we can calculate $\Omega_D h^2$ \cite{density1, density2, density3}by
\begin{eqnarray}
 &&\Omega_D h^2\simeq{\frac{1.07\times10^9 x_F}{\sqrt{g_*} M_{pl} (a+3b/x_F)\rm{GeV}}}.
\label{relic density}
\end{eqnarray}
\subsection{$a$ and $b$ of $\chi^0_L$}
 We give the most important lepton neutralino $(\chi^0_L)$ annihilation diagrams whose final states are leptons and light neutrinos. For a complete list of all tree level processes(in FIG.1), we can calculate $a$ and $b$ (in the low velocity limit) by using the couplings in Eqs.(\ref{coupling1}-\ref{coupling3}).
\begin{figure}[t]
\centering
\includegraphics[width=16cm]{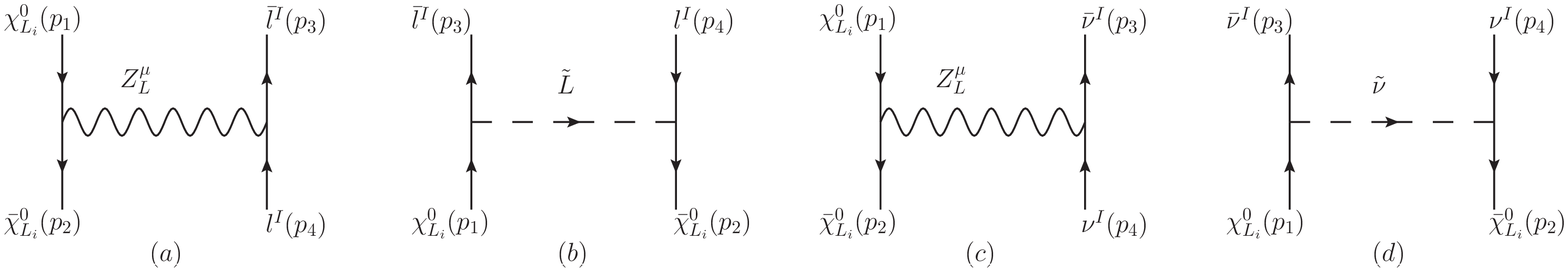}\\
\caption{Feynman diagrams for the $\chi^0_{L_i}\chi^0_{L_i}\rightarrow Z^{\mu}_L\rightarrow\bar{l}^Il^I(\bar{\nu}^I\nu^I)$ and $\chi^0_{L_i}\chi^0_{L_i}\rightarrow \tilde{L}(\tilde{\nu})\rightarrow\bar{l}^Il^I(\bar{\nu}^I\nu^I)$ at the tree-level.}
\end{figure}
\begin{figure}[t]
\centering
\includegraphics[width=16cm]{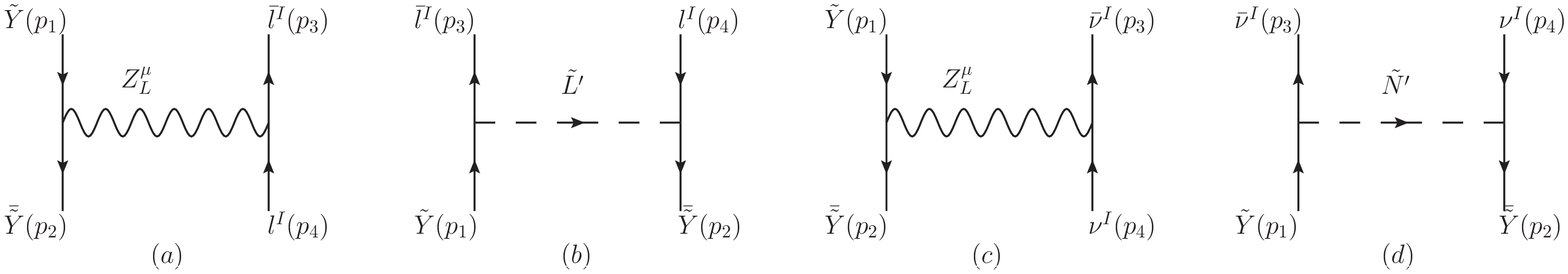}\\
\caption{Feynman diagrams for the $\tilde{Y}\tilde{Y}\rightarrow Z^{\mu}_L\rightarrow\bar{l}^Il^I(\bar{\nu}^I\nu^I)$ and $\tilde{Y}\tilde{Y}\rightarrow \tilde{L'}(\tilde{N'})\rightarrow\bar{l}^Il^I(\bar{\nu}^I\nu^I)$ at the tree-level.}
\end{figure}
\begin{eqnarray}
 &&X_1=Z^{5i*}_{N_L}Z^{5i}_{N_L}-Z^{4i*}_{N_L}Z^{4i}_{N_L},
  \qquad \quad X_2=Z^{1i}_{N_L}Z^{Ij}_{\tilde{L}}+Z^{1i*}_{N_L}Z^{(I+3)j}_{\tilde{L}},
\nonumber\\&& X_3=Z^{1i}_{N_L}Z^{Ij}_{\tilde{L}}-Z^{1i*}_{N_L}Z^{(I+3)j}_{\tilde{L}},
\qquad X^{'}_1=\frac{1}{\sqrt{2}}X_1,
\qquad \quad
\nonumber\\&&  X^{'}_2=X^{'}_3=Z^{1i}_{N_L}Z^{Jj*}_{\tilde{\nu}},
 \qquad \qquad \quad    A_1=|X_2|^2+|X_3|^2,
\nonumber\\&&   A_2=-|X_2|^2+|X_3|^2,
\qquad  \qquad  \quad    A_3=|X_2|^2-|X_3|^2,
\nonumber\\&&A_4=X_2 X^{*}_3+X^{*}_2 X_3,
\qquad \qquad \quad A^{'}_1=|X^{'}_2|^2+|X^{'}_3|^2=2|X^{'}_2|^2,
\nonumber\\&&A^{'}_2=A^{'}_3=0,
\qquad \qquad  \qquad  \qquad  \  A^{'}_4=X^{'}_2X^{'*}_3+X^{'*}_2X^{'}_3=2|X^{'}_2|^2.
\label{Xjian}
\end{eqnarray}

We define above terms to simplify the following formulas. They are all expressions related to couplings. Next we write the concrete expression of $a$ and $b$.

\begin{eqnarray}
&&a=\sum_{l=e,\mu,\tau}\Big\{\frac{g^4_L}{8\pi}(1+\frac{m^2_l}{m^2_D})^{\frac{1}{2}}\Big[\frac{9(2m^2_D+m^2_l)}{(4m^2_D-m^2_{Z_L})^2} |X_1|^2+\frac{(m_D A_1+m_l A_3)^2}{4 (m^2_D-m^2_l+m^2_{\tilde{L}})^2}
\nonumber\\&&\hspace{0.8cm}
+\sum^6_{j=1}\frac{3(2m^2_D+m^2_l)A_1-3m_D m_l A_2+6 m_D m_l A_3}{2(4m^2_D-m^2_{Z_L})(m^2_D-m^2_l+m^2_{\tilde{L_j}})}X^{*}_1\Big] \Big\}
\nonumber\\
&&\hspace{0.8cm}+\sum_{\chi^0_{N_\alpha}=\nu_e,\nu_\mu,\nu_\tau}\Big\{\frac{g^4_L m^2_D}{8\pi}\Big[\frac{18}{(4m^2_D-m^2_{Z_L})^2} |X^{'}_1|^2
+\frac{1}{(m^2_D+m^2_{\tilde{\nu}})^2} |X^{'}_2|^4\nonumber\\
&&\hspace{0.8cm}-\frac{6\sqrt{2}}{(4m^2_D-m^2_{Z_L})(m^2_D+m^2_{\tilde{\nu}})} X^{'*}_1 |X^{'}_3|^2\Big]\Big\},
\label{Xa}
\end{eqnarray}
\begin{eqnarray}
&&b=\sum_{l=e,\mu,\tau}\Big\{\frac{g^4_L}{16\pi}\Big[(\frac{18m^2_D}{(4m^2_D-m^2_{Z_L})^2}+\frac{12m^4_D-21m^2_D m^2_{Z_L}}{(4m^2_D-m^2_{Z_L})^3})|X_1|^2+(\frac{m^2_D}{4(m^2_D+m^2_{\tilde{L}})^2}
 \nonumber\\
&&\hspace{0.8cm}+\frac{15m^6_D+10m^4_Dm^2_{\tilde{L}}+7m^2_Dm^4_{\tilde{L}}}{12(m^2_D+m^2_{\tilde{L}})^4})A^2_1+\sqrt{2}X^{*}_1 A_2(\frac{3m_D m_l}{2(4m^2_D-m^2_{Z_L}) (m^2_D+m^2_{\tilde{L}})}
 \nonumber\\
&&\hspace{0.8cm}+\frac{-3m^5_D m_l m^2_{Z_L}-16m^5_D m_l m^2_{\tilde{L}}-2m^3_D m_l m^2_{Z_L} m^2_{\tilde{L}}-3m_D m_l m^2_{Z_L} m^4_{\tilde{L}}}{4(4m^2_D-m^2_{Z_L})^2 (m^2_D+m^2_{\tilde{L}})^3})
\nonumber\\
&&\hspace{0.8cm}+(\frac{m^2_D(-5A_1+2A_4)-m_D m_l A_3}{2(4m^2_D-m^2_{Z_L}) (m^2_D+m^2_{\tilde{L}})}+\frac{-m^2_D(m^2_D(A_1-A_4)-m_D m_l A_3)}{(4m^2_D-m^2_{Z_L}) (m^2_D+m^2_{\tilde{L}})^2})\sqrt{2}X^{*}_1
\nonumber\\
&&\hspace{0.8cm}+(\frac{7m^5_D m_l-2m^3_D m_l m^2_{\tilde{L}}+3m_D m_l m^4_{\tilde{L}}}{12(m^2_D+m^2_{\tilde{L}})^4}+\frac{m_D m_l}{2(m^2_D+m^2_{\tilde{L}})^2})A_1 A_3
 \nonumber\\
&&\hspace{0.8cm}+\sqrt{2}(m^2_D X^{*}_1 A_1+m_D m_l X^{*}_1 A_3)(-\frac{3m_D m_l}{(4m^2_D-m^2_{Z_L}) (m^2_D+m^2_{\tilde{L}})}
\nonumber\\
&&\hspace{0.8cm}+\frac{m^2_D(10m^4_D-m^2_D m^2_{Z_L}+24m^2_D m^2_{\tilde{L}}-3m^2_{Z_L} m^2_{\tilde{L}}+6m^4_{\tilde{L}})}{(4m^2_D-m^2_{Z_L})^2 (m^2_D+m^2_{\tilde{L}})^3})\Big]\Big\}
\nonumber\\
&&\hspace{0.8cm}+\sum_{\chi^0_{N_\alpha}=\nu_e,\nu_\mu,\nu_\tau}\Big\{(m_l\rightarrow m_{\nu}\simeq0, m_{\tilde{L}}\rightarrow m_{\tilde{\nu}},X_1\rightarrow X^{'}_1, A_k\rightarrow A^{'}_k(k=1,2,3,4)\Big\}.
\label{Xb}
\end{eqnarray}

Because the mass of the light neutrino is too small, we regard it as zero. When we compute the term $b$, we find the specific form of $b$ is tedious and complicated. To simplify the results, we perform Taylor expansion on $\frac{m_l}{m_D}$ and retain it to the second order.
\subsection{$a$ and $b$ of $\tilde{Y}$}
Similarly, we can also calculate the results of $a$ and $b$ for $\tilde{Y}$. We give the dominant contribution to the annihilation cross section come from $\bar{l}^Il^I$ and $\bar{\nu}^I\nu^I$. The tree diagrams are shown in FIG.2. To simplify the results, we use the following assumptions:
 \begin{eqnarray}
  &&Y_1=2+L_4, \qquad \qquad \qquad \qquad Y_2=\lambda_4 Z^{4i*}_{\tilde{L'}}-\lambda_6 Z^{3i*}_{\tilde{L'}},
\nonumber\\&&  Y_3=-\lambda_4 Z^{4i*}_{\tilde{E}}-\lambda_6 Z^{3i*}_{\tilde{E}},\quad \qquad Y^{'}_1=Y_1,
  \nonumber\\&& Y^{'}_2=\lambda_4 Z^{4i*}_{\tilde{N'}},\qquad \qquad \quad \qquad \ \ B_1=|Y_2|^2+|Y_3|^2,
 \nonumber\\&&  B_2=-|Y_2|^2+|Y_3|^2, \qquad \qquad \ B_3=|Y_2|^2-|Y_3|^2.
\label{Yjian}
 \end{eqnarray}

Using the couplings in Eqs.(\ref{coupling4}-\ref{coupling6}) and the expressions in Eq.(\ref{Yjian}), we deduce the results of $a$ and $b$.

  \begin{eqnarray}
  && a=\sum_{l=e,\mu,\tau}\Big\{\frac{g^2_L}{8\pi}(1+\frac{m^2_l}{m^2_D})^{\frac{1}{2}}\Big[\frac{4g^2_L(2m^2_D+m^2_l)}{(4m^2_D-m^2_{Z_L})^2} |Y_1|^2+\frac{(m_D B_1+m_lB_3)^2}{16 g^2_L(m^2_D-m^2_l+m^2_{\tilde{L'}})^2}
\nonumber\\
&&\hspace{0.8cm}+\frac{(2m^2_D+m^2_l)B_1-m_D m_l B_2+2m_D m_l B_3}{2(4m^2_D-m^2_{Z_L})(m^2_D-m^2_l+m^2_{\tilde{L'}})}Y^{*}_1\Big]\Big\}\nonumber\\
&&\hspace{0.8cm}+\sum_{\chi^0_{N_\alpha}=\nu_e,\nu_\mu,\nu_\tau}\Big\{\frac{g^2_L m^2_D}{8\pi}\Big[\frac{4g^2_L}{(4m^2_D-m^2_{Z_L})^2} |Y^{'}_1|^2+\frac{1}{4 g^2_L(m^2_D+m^2_{\tilde{N'}})^2}|Y^{'}_2|^4\Big]\Big\},
 \label{Ya}
 \end{eqnarray}
 \begin{eqnarray}\nonumber\\&&b=\sum_{l=e,\mu,\tau}\Big\{\frac{g^2_L}{16\pi}\Big[(\frac{8m^2_D g^2_L}{(4m^2_D-m^2_{Z_L})^2}+\frac{4(-2m^4_D-5m^2_D m^2_{Z_L})g^2_L}{3(4m^2_D-m^2_{Z_L})^3}) |Y_1|^2+(\frac{m^2_D}{4g^2_L(m^2_D+m^2_{\tilde{L'}})^2}
\nonumber\\
&&\hspace{0.8cm}+\frac{15m^6_D+10m^4_Dm^2_{\tilde{L'}}+7m^2_Dm^4_{\tilde{L'}}}{12g^2_L(m^2_D+m^2_{\tilde{L'}})^4})B^2_1+B_1B_3(\frac{7m^5_D m_l-2m^3_D m_l m^2_{\tilde{L'}}+3m_D m_l m^4_{\tilde{L'}}}{12g^2_L(m^2_D+m^2_{\tilde{L'}})^4}
\nonumber\\
&&\hspace{0.8cm}+\frac{m_D m_l}{2g^2_L(m^2_D+m^2_{\tilde{L'}})^2})+(\frac{m^2_D(10m^4_D-m^2_D m^2_{Z_L}+24m^2_D m^2_{\tilde{L'}}-3m^2_{Z_L} m^2_{\tilde{L'}}+6m^4_{\tilde{L'}})}{3(4m^2_D-m^2_{Z_L})^2 (m^2_D+m^2_{\tilde{L'}})^3}
\nonumber\\
&&\hspace{0.8cm}-\frac{1}{(4m^2_D-m^2_{Z_L}) (m^2_D+m^2_{\tilde{L'}})})(m^2_D B_1+m_D m_l B_3)+(\frac{-5m^2_DB_1-3m_D m_l B_3}{6(4m^2_D-m^2_{Z_L}) (m^2_D+m^2_{\tilde{L'}})}
\nonumber\\
&&\hspace{0.8cm}-\frac{m^4_DB_1-m^2_DB_3}{3(4m^2_D-m^2_{Z_L}) (m^2_D+m^2_{\tilde{L'}})^2})Y^{*}_1+Y^{*}_1 B_2(\frac{m_D m_l}{2(4m^2_D-m^2_{Z_L}) (m^2_D+m^2_{\tilde{L'}})}
\nonumber\\
&&\hspace{0.8cm}+\frac{-3m^5_D m_l m^2_{Z_L}-16m^5_D m_l m^2_{\tilde{L'}}-2m^3_D m_l m^2_{Z_L} m^2_{\tilde{L'}}-3m_D m_l m^2_{Z_L} m^4_{\tilde{L'}}}{12(4m^2_D-m^2_{Z_L})^2 (m^2_D+m^2_{\tilde{L'}})^3})\Big]\Big\}\nonumber\\
&&\hspace{0.8cm}+\sum_{\chi^0_{N_\alpha}=\nu_e,\nu_\mu,\nu_\tau}\Big\{\frac{m^2_D}{16\pi}\Big[(\frac{15m^4_D+10m^2_Dm^2_{\tilde{N'}}+7m^4_{\tilde{N'}}}{12(m^2_D+m^2_{\tilde{N'}})^4}+
\frac{m^2_D}{4(m^2_D+m^2_{\tilde{N'}})})|Y^{'}_2|^4
\nonumber\\
&&\hspace{0.8cm}+(\frac{4g^2_L}{(4m^2_D-m^2_{Z_L})^2}+
\frac{g^2_L(-4m^2_D-5m^2_{Z_L})}{3(4m^2_D-m^2_{Z_L})^3})Y^{'*}_1\Big]\Big\}.
\label{Yb}
\end{eqnarray}
The mass rotation matrices corresponding to $\chi^0_L$, $\tilde{L}$, $\tilde{\nu}$, $\tilde{L'}$ and $\tilde{N'}$ are $Z_{N_L}$, $Z_{\tilde{L}}$, $Z_{\tilde{\nu}}$, $Z_{\tilde{L'}}$ and $Z_{\tilde{N'}}$.
\section{Numerical results}
We are now in a position to present some numerical results. Current data imply that dark mater is five times more prevalent than normal matter and accounts for about a quarter of the universe. In section I, we give precisely the constrain of the relic density of cold non-baryonic dark matter and it is $\Omega_D h^2=0.1186\pm0.0020$\cite{numerical}. Next we will discuss $\Omega h^2$ of the $\chi^0_L$ and $\tilde{Y}$.
\begin{figure}[t]
\begin{center}
\begin{minipage}[c]{0.48\textwidth}
\includegraphics[width=2.9in]{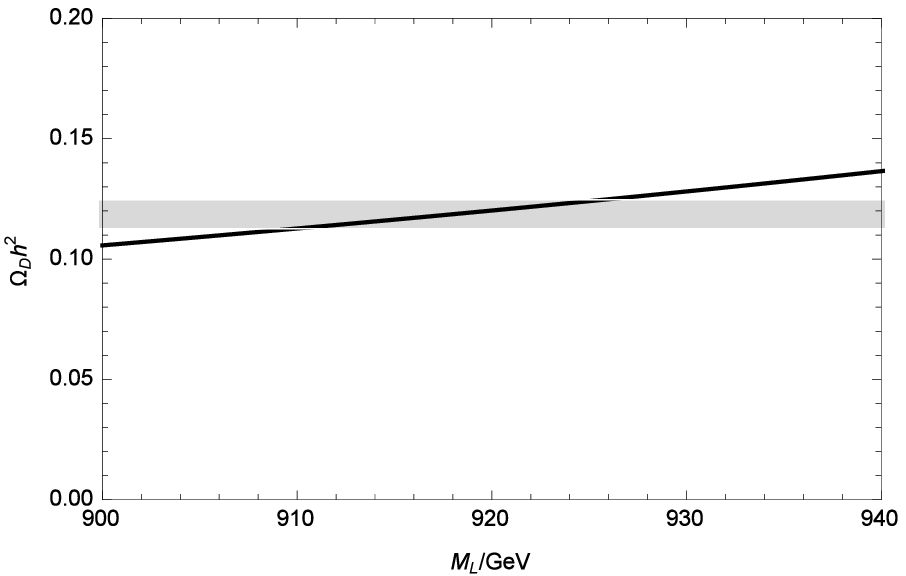}
\end{minipage}%
\begin{minipage}[c]{0.45\textwidth}
\includegraphics[width=2.9in]{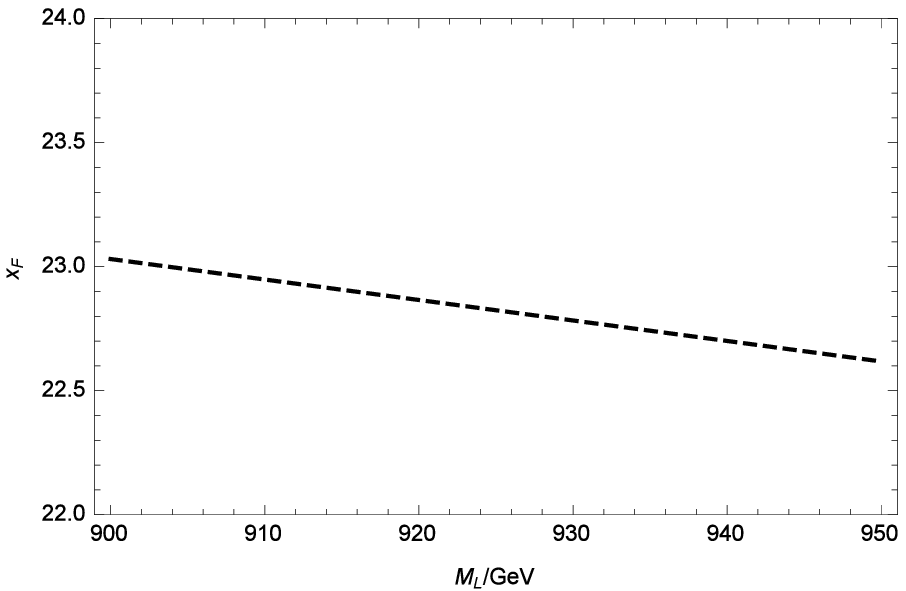}
\end{minipage}
\caption{The relic density and $x_F$ versus $M_L$ for $\chi^0_L$.} \label{fig3}
\end{center}
\end{figure}
\begin{figure}[t]
\begin{center}
\begin{minipage}[c]{0.48\textwidth}
\includegraphics[width=2.9in]{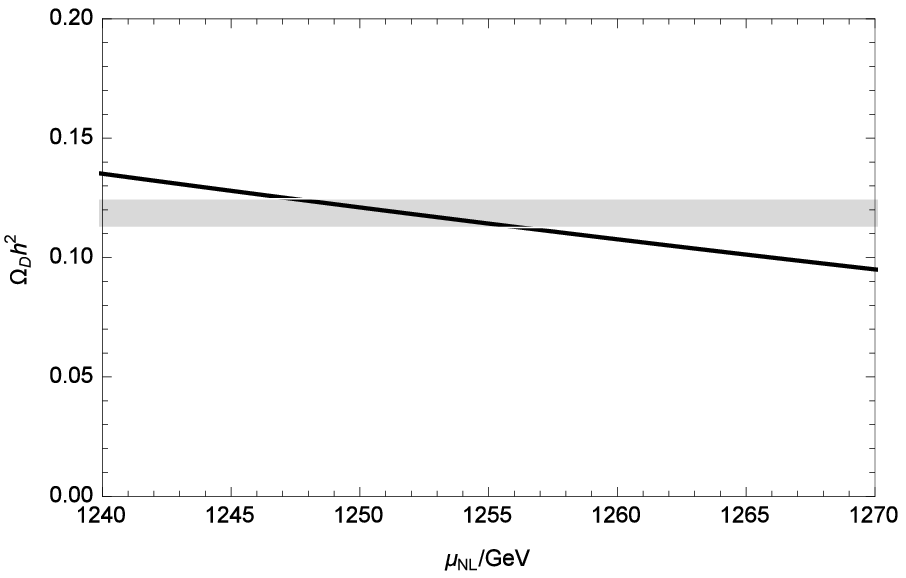}
\end{minipage}%
\begin{minipage}[c]{0.45\textwidth}
\includegraphics[width=2.9in]{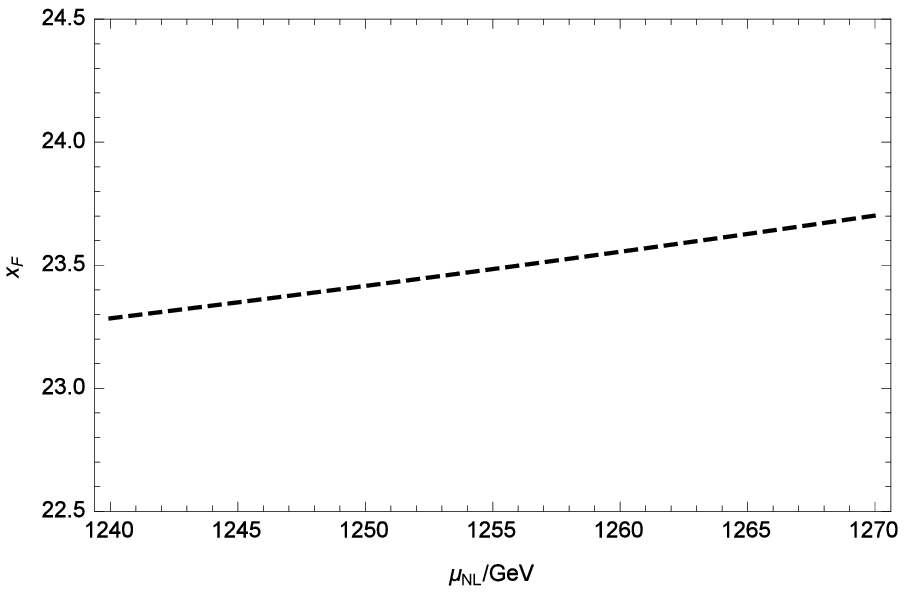}
\end{minipage}
\caption{The relic density and $x_F$ versus $\mu_{NL}$ for $\chi^0_L$.} \label{fig4}
\end{center}
\end{figure}
\begin{figure}[t]
\begin{center}
\begin{minipage}[c]{0.48\textwidth}
\includegraphics[width=2.9in]{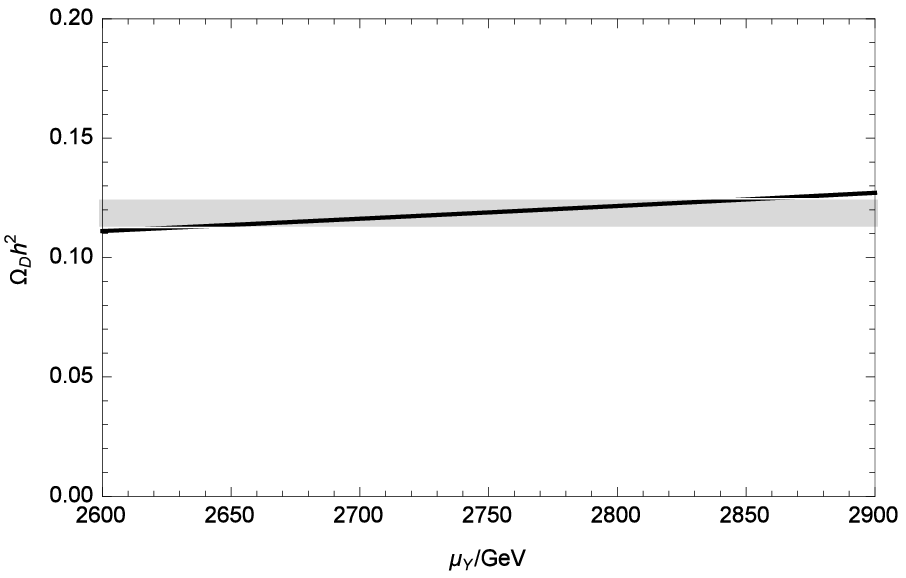}
\end{minipage}%
\begin{minipage}[c]{0.48\textwidth}
\includegraphics[width=2.9in]{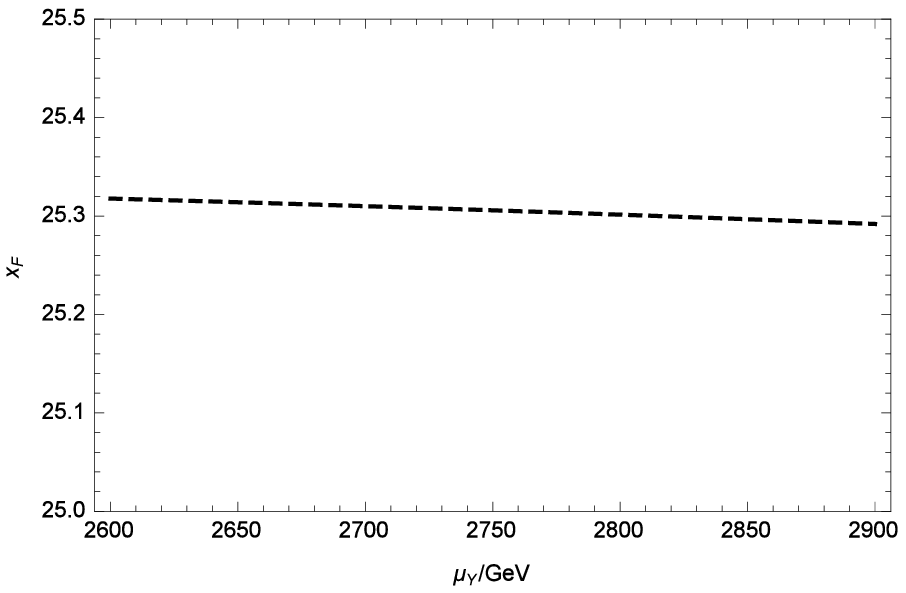}
\end{minipage}
\caption{The relic density and $x_F$ versus $\mu_{Y}$ for $\tilde{Y}$.} \label{fig5}
\end{center}
\end{figure}
\begin{figure}[t]
\begin{center}
\begin{minipage}[c]{0.48\textwidth}
\includegraphics[width=2.9in]{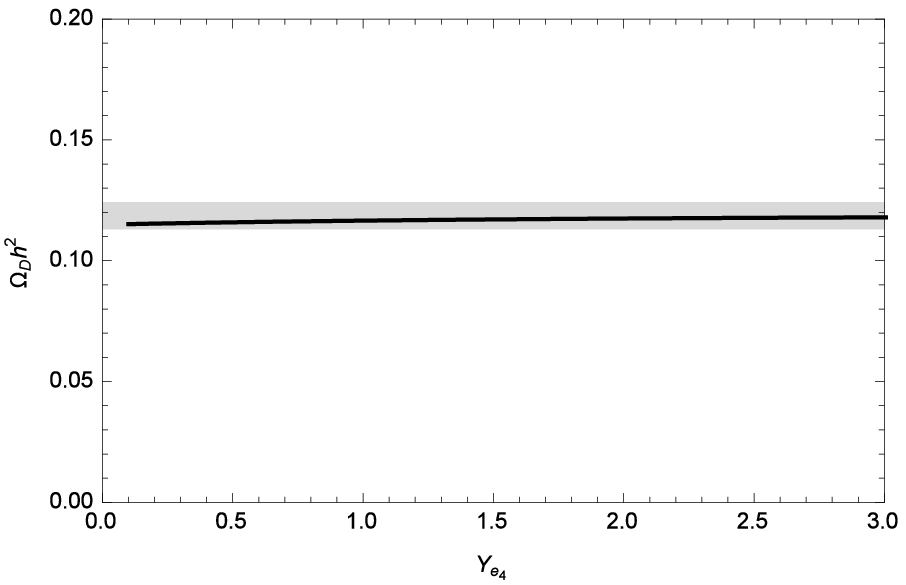}
\end{minipage}%
\begin{minipage}[c]{0.48\textwidth}
\includegraphics[width=2.9in]{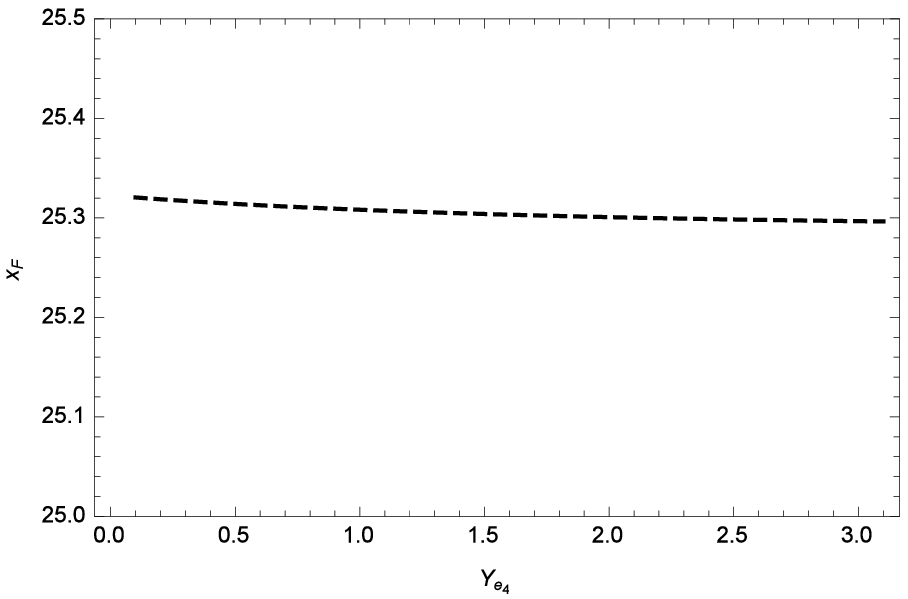}
\end{minipage}
\caption{The relic density and $x_F$ versus $Y_{e4}$ for $\tilde{Y}$.} \label{fig6}
\end{center}
\end{figure}
\subsection{Numerical result of $\chi^0_L$}
To obtain a more transparent numerical results, we adopt the following assumptions on parameter space:
\begin{eqnarray}
&& g_L=1/6,\ \ \  \mu=0.5{\rm TeV},\ \ \  \upsilon_{Nlt}=3{\rm TeV}, \ \ \ \tan{\beta}=10, \ \ \ (\lambda_{N^c})_{ii}=1,
\nonumber\\
&&\hspace{0cm} \tan{\beta_L}=\tan{\beta_{NL}}=2, \ \ \  (A_l)_{ij}=(A_{l^{'}})_{ij}=0(i\neq j), \ \ \ (A_l)_{ii}=2{\rm TeV},
\nonumber\\
&&\hspace{0cm} (A_{l^{'}})_{ii}=0.3 {\rm TeV}, \ \ \  (M^2_{\tilde{L}})_{ii}=4{\rm TeV^2}, \ \ \ (M^2_{\tilde{L}})_{ij}=0.1{\rm TeV^2}(i \neq j).
\label{Xcanshu}
\end{eqnarray}

Here $i$=1,2,3, $\mu_L$=$\mu_{NL}$=0.8TeV. The lightest $m_{\chi^0_L}$ mass is denoted by $m_D$, $m_{\chi^0_L}$=$m_D$. We see from FIG.3 that in order to constrain the region of parameter space in EBLMSSM we need satisfy experimental results of $\Omega_D h^2$. We study relic density $\Omega_D h^2$ and $x_F$ versus $M_L$. The grey area is the experimental results of $\Omega_D h^2$ in $3\sigma$ and the solid line represents relationship between $\Omega_D h^2$ and $M_L$. Parameter $M_L$ presents in the diagonal parts of the $\chi^0_L$ mass matrix. Because the increase of $M_L$ leads to the decrease of $\langle\sigma v\rangle$ and the decrease of $\langle\sigma v\rangle$ causes $\Omega_D h^2$ to increase, $\Omega_D h^2$ increases with the increase of $M_L$. Furthermore, parameter $M_L$ has a very large impact on $\Omega_D h^2$. As sensitive parameter $M_L$ is limited to 908-929GeV, when $\Omega_D h^2$ satisfies the experiment bounds. At this time, the mass of $m_D$ is limited to about 300GeV. On the other hand, the range of variation is very small for $x_F$, about 0.5. The curve of  $x_F$ increases as $M_L$ decreases, which is a negative correlation.

The strong impact of model parameters $\mu_{NL}$ on the $\Omega_D h^2$ is further illustrated by FIG.4. $\mu_{NL}$ is related to the non-diagonal parts of the $\chi^0_L$ mass matrix in the EBLMSSM. In addition to the parameters of Eq.(\ref{Xcanshu}), we also let $M_L$=2TeV and $\mu_L$=1.1TeV. As can be seen from FIG.4, when $\mu_{NL}$ increases, $\Omega_D h^2$ shows a downward trend. Different from the diagonal element $M_L$ is that $\langle\sigma v\rangle$ increases as $\mu_{NL}$ increases. As sensitive parameters $\mu_{NL}$ is limited to 1248-1256GeV when $\Omega_D h^2$ satisfies the experiment bounds. It's range is about half of the parameter $M_L$. Besides, the curve of $x_F$ is very slowly rising. $x_F$ is related to $\langle\sigma v\rangle$ and increases as $\langle\sigma v\rangle$ increases. This is the reason for the decrease in $\Omega_D h^2$.

\subsection{Numerical result of $\tilde{Y}$}
To obtain the numerical results, we adopt the following parameters as
\begin{eqnarray}
&&L_4=1.5,\ \ \ Y_{e5}=1.2,\ \ \ \mu=0.5{\rm TeV} ,\ \ \ Y_{\nu4}=Y_{\nu5}=1.2,
\nonumber\\
&&\hspace{0cm} \mu_{NL}=1{\rm TeV},\ \ \  \lambda_E=\lambda_L=\lambda_{NL}=\lambda^{'},\ \ \ \lambda_4=\lambda_5=\lambda_6=L_m,
\nonumber\\
&&\hspace{0cm}A_{e_4}=A_{e_5}=A_{\nu_4}=A_{\nu_5}=1{\rm TeV},\ \ \ \ A_{LL}=A_{LE}=A_{LN}=1{\rm TeV},
\nonumber\\
&&\hspace{0cm}M^2_{\tilde{L}_4}=M^2_{\tilde{L}_5}=M^2_{\tilde{e}_4}=M^2_{\tilde{e}_5}=M^2_{\tilde{\nu}_4}=M^2_{\tilde{\nu}_5}=1\rm TeV^2.
\label{Ycanshu}
\end{eqnarray}

The parameters that are repeated with the subsection A are not listed, they can all be found in Eq.(\ref{Xcanshu}).
In FIG.5, we study effects from the new parameters $\mu_{Y}$ on our numerical results. Actually $\mu_{Y}$=$m_D$, $\lambda^{'}$=0.7, $Y_{e4}$=1.3, $L_m$=0.8. The reasonable range of $\mu_{Y}$ is 2650-2850GeV. The FIG.5 shows that larger $\mu_{Y}$ can lead to larger $\Omega_D h^2$, $\langle\sigma v\rangle$ decreases at the same time. The curve trend of $x_F$ is the opposite of $\Omega_D h^2$.

In FIG.6, taking $\lambda^{'}$=1, $L_m$=0.88, $\mu_{Y}$=$m_D$=2700GeV, we plot the $\Omega_D h^2$ and $x_F$ versus $Y_{e4}$. $Y_{e4}$ is the Yukawa coupling constant that can influence the mass matrix of exotic slepton. When $Y_{e4}$ gradually becomes larger, $\Omega_D h^2$ also increases accordingly. And $x_F$ is limited to 24.6-24.8, almost no change within the scope of meeting the relic density boundaries of dark matter. In general, the changes in both the dashed line and the solid line are flat and stable. The reason is that $m_D$ is taken as a fixed value in this figure.

\section{Discussion and conclusion}
The extension of the BLMSSM model by the addition of exotic superfields $\phi_{NL}$, $\varphi_{NL}$, $Y$ and $Y'$ which can make the exotic leptons heavy and unstable. We give it a new name called EBLMSSM, where we can deduce the mass matrices of particles and the couplings. The spinor $\tilde{Y}$ and the mixing of superfields $Y$, $Y^{'}$ are all new terms beyond BLMSSM. And the exotic slepton($\tilde{L}^{'}$) and exotic sneutrino($\tilde{N}^{'}$) of generations 4 and 5 mix. The above enriches the content of new physics and dark matter physics.

We choose the lightest $\chi^0_L$ and $\tilde{Y}$ as dark matter candidates due to that they are consistent with the characteristics of cold dark matter. Then we research the relic density of $\chi^0_L$ and $\tilde{Y}$. In rational parameter space, $\Omega_D h^2$ can match the experiment bounds. And based on experimental data we can give confine on sensitive parameters. The EBLMSSM inevitably will be a feature of many particle physics models beyond the standard model. We believe the results presented here may generally be useful in the study of such models and of their cosmological consequences.

{\bf Acknowledgments}

We are very grateful to Wei Chao the teacher of Beijing Normal University, for giving us some useful discussions. This work is supported by National Natural Science Foundation of China (NNSFC) (No. 11535002, No. 11605037, No. 11705045), the Natural Science Foundation of Hebei province with Grant
No. A2016201010 and No. A2016201069, Hebei Key Lab of Optic-Electronic Information and Materials, the midwest universities comprehensive strength promotion project and the youth top-notch talent support program of the Hebei Province.

\end{document}